\documentclass[epj,nopacs]{svjour}
\usepackage{amsmath}
\usepackage{amssymb}
\usepackage{latexsym}
\usepackage[dvips]{graphicx}
\usepackage[english]{babel}

\def\bge{\begin{equation}}
\def\ene{\end{equation}}
\def\bgea{\begin{eqnarray}}
\def\enea{\end{eqnarray}}

\def\bge{\begin{equation}}
\def\ene{\end{equation}}
\def\bgea{\begin{eqnarray}}
\def\enea{\end{eqnarray}}

\def\ls{\raise 1.5pt\hbox{$\,<\;$}\kern -10.5pt\lower3.5pt
          \hbox{$\sim$}\kern 1.5pt} 
\def\gs{\raise 1.5pt\hbox{$\,>\,$}\kern -9.5pt\lower3.5pt
          \hbox{$\sim$}\kern 1.5pt} 

\usepackage{color}

\usepackage{ulem} 

\begin{document}
\sloppy
\title{Revisiting the cosmic distance duality relation with machine learning
reconstruction methods: the combination of HII galaxies and
ultra-compact radio quasars}
\author{Tonghua Liu$^{1,2}$, Shuo Cao$^{2,3\ast}$, Sixuan Zhang$^{4}$, Xiaolong Gong$^{1}$, Wuzheng Guo$^{2}$, and Chenfa Zheng$^{2}$}
\institute{ $^1$School of Physics and Optoelectronic, Yangtze
University, Jingzhou 434023, China;\\
$^2$Department of Astronomy, Beijing Normal University, Beijing 100875, China; \email{caoshuo@bnu.edu.cn}\\
$^3$Advanced Institute of Natural Sciences, Beijing Normal
University at Zhuhai 519087, China\\
$^4$ Graduate School of Advanced Science and Engineering, Hiroshima University, Hiroshima 739-8526, Japan}

\authorrunning{Liu et al.}
\titlerunning{Testing CDDR with GP and ANN}

\date{May 6, 2021}

\abstract{In this paper, we carry out an assessment of cosmic
distance duality relation (CDDR) based on the latest observations of
HII galaxies acting as standard candles and ultra-compact structure
in radio quasars acting as standard rulers. Particularly, two
machine learning reconstruction methods (Gaussian Process (GP) and
Artificial Neural Network (ANN)) are applied to reconstruct the
Hubble diagrams from observational data. We show that both
approaches are capable of reconstructing the current constraints on
possible deviations from the CDDR in the redshift range $z\sim
2.3$. Considering four different parametric methods of CDDR, which
quantify deviations from the CDDR and the standard cosmological
model, we compare the results of the two different machine learning
approaches. It is observed that the validity of CDDR is in well
agreement with the current observational data within $1\sigma$ based
on the reconstructed distances through GP in the overlapping
redshift domain. Moreover, we find that ultra-compact radio quasars
could provide $10^{-3}$-level constraints on the violation parameter
at high redshifts, when combined with the observations of HII
galaxies. In the framework of ANN, one could derive robust
constraints on the violation parameter at a precision of $10^{-2}$,
with the validity of such distance duality relation within $2\sigma$
confidence level.}

\maketitle

\section{Introduction}

The cosmic distance duality relation (CDDR) is a fundamental
relation in modern cosmology, which relates two cosmologiccal
distances in cosmology (i.e. the luminosity distance $D_L(z)$ and
angular diameter distance $D_A(z)$). More specifically, the CDDR
indicates that $D_L(z)$ and $D_A(z)$ satisfy the relation of
$D_L(z)=D_A(z)(1+z)^2$ at the same redshift \cite{1,2}.
Theoretically, the validity of the CDDR depends on three basic
assumptions: i)the space-time is described by a metric theory;
ii)the light travels along the null geodesics between the source and
the observer; iii)the photon number is conserved. Moreover, one of
the basic assumptions of general relativity is that photons travel
along null geodesics. In other word, the validity of the CDDR can be
a support of general relativity in some extents. As a fundamental
relation, the CDDR has been widely used in varieties of research
fields in astronomy, such as the large-scale distribution of
galaxies and the near-uniformity of the CMB temperature \cite{3}, as
well as the gas mass density profile and temperature profile of
galaxy clusters \cite{Cao11a,Cao16}. Various astrophysical
mechanisms, such as gravitational lensing, and dust extinction, may
cause the deviation of the CDDR from the view of observation. More
specifically, photons emitted from the source are affected in the
process of propagation due to gravitational lensing effects and dust
extinctions. Consequently, the necessary conditions for maintaining
the CDDR are violated. Therefore, it is necessary to test the
reliability of the CDDR accurately before applying to various
astronomical theories.

Traditionally, testing CDDR needs two types of observational data
sets, i.e., the luminosity distance derived from the luminous
sources with known (or standardizable) intrinsic luminosity in the
Universe like type-Ia supernova (SN Ia), and the angular diameter
distance observed from Baryon Acoustic Oscillations (BAO) \cite{4},
Sunyaev-Zeldovich (SZ) effect in clusters with X-ray surface
luminosity measurements \cite{5,Cao11b,6}, or strong gravitational
lensing (SGL) \cite{7,8}, etc. However, it is necessary to point out that
the luminosity distance inferred from SN Ia only covers the relatively lower redshift
range $z\leq 1.4$. The so-called "nuisance" parameter of SN Ia
usually optimizes along with model parameters in the chosen
cosmological model \cite{9}. Meanwhile, the angular diameter
distance derived from BAO or SZ effect is strongly model-dependent,
thus will bring systematic uncertainties which are hard to quantify
and affect the validity of testing CDDR. In addition, other works
\cite{10,11,12} attempted to apply the BAO observations to CDDR test,
which also suffers from the limited sample size and low
redshift range $0.35\leq z\leq 0.74$. Therefore, in order to perform
the validity of testing CDDR, one needs to reduce the statistical
uncertainty by increasing the depth and quality of the observed data
set. Meanwhile, the redshift ranges of the two samples that inferred
the angular diameter distance and the luminosity distance should be
roughly consistent. Such issue has been recently discussed in Ref.~\cite{Zheng21}, focusing on a new idea of
testing CDDR through the multiple measurements of high-redshift quasars.

Although many efforts have been made to perform robust tests of
CDDR, the lack of adequate observational samples and
model-independent methods should be taken into account. Specially,
it is difficult to obtain samples that satisfy both the luminosity
distance and the angular diameter distance in roughly the same
redshift range. This redshift-matching problem problem was
recognized a long time ago \cite{Cao11b}, with a heuristic
suggestion that the choice of redshfit difference $\Delta z$ could
play an important role in model-independent tests of such relation.
More recently, many authors  presented a new way to constrain the
CDDR with different machine learning algorithms \cite{12a,12b,12d},
with the luminosity distance and angular diameter distance
reconstructed from complementary external probes (Type Ia supernovae
and gravitational wave (GW) standard sirens) \cite{12c}. Their
results demonstrated the effectiveness of machine learning
approaches in the high-precision test of the electromagnetic and
gravitational distance duality relations. More importantly,
considering the fact that the purpose of modern cosmology is to
establish consistent and robust theories, all alternative methods of
testing the fundamental principles of cosmology are necessary. In
this paper, we will use two non-parameterized methods, Gaussian
Process (GP) and Artificial Neural Network (ANN) algorithm, to
reconstruct the newest observations of HII galaxy Hubble diagram and
ultra-compact structure of radio quasars, respectively. These two
approaches are data-driven and have no assumptions about the data,
suggesting that they are completely model-independent. The
luminosity distance is inferred from reconstructed HII galaxy Hubble
diagram and the angular diameter distance is obtained from the
angular-size relation of compact radio quasar. The advantage of
using these two data is that the redshift ranges of the two samples
are roughly consistent, and can reach a relatively high redshift
range $z\sim 2.33$. Since no models were assumed in our analysis,
our method produced a clear measurement on the CDDR.

This paper is organized as follows: in Section 2 we briefly
introduce methodology of deriving two different cosmological
distances from the HII galaxies (including extragalactic HII
regions) and ultra-compact structure of radio quasar sources,
respectively. The two non-parameterized methods, GP and ANN
reconstructing the two data sets are described in Section 3. In
Section 4, we show the methodology of measuring CDDR and show our
results. Finally, we summarize our conclusions in Section 5.

\section{Data and Methodology}\label{sec:data}

\subsection{Luminosity distances from HII galaxies and extragalactic HII regions}

In order to measure luminosity distances in the Universe, we always
turn to sources that have known (or standardised) intrinsic
luminosity, such as type-Ia supernova (SN Ia) \cite{9}, more distant
quasars \cite{13,14,15,16}, and gamma-ray bursts (GRB) \cite{17},
etc. In addition, the HII galaxies and extragalactic HII regions
\cite{18,19,20} constitute a large fraction of population that can
be observed up to very high redshifts, beyond the feasible limits of
supernova studies. It is well known that the luminosity
$L($H$\beta)$ in H$\beta$ and the ionized gas velocity dispersion
$\sigma$ of HII galaxies and extragalactic HII regions may have a
quantitative relation (be known as ``$L$--$\sigma$" relation). The
physics behind this relation is based only on a simple idea, i.e.,
as the mass of the starburst component increases, the number of
ionized photons and the turbulent velocity of the gas may both
increase as well. Melnick \textit{et al.} first found that the
scatter of ``$L$--$\sigma$" relation is very small and have the
capability to determine the cosmological distance independent of
redshift\cite{21}. More specifically, based on the measured flux
density (or luminosity) and the turbulent velocity of the gas, one
can infer the luminosity distance directly.  Whereafter, the
validity of the ``$L$--$\sigma$" relation acting as the standard
candle and its possible cosmological applications have been
extensively discussed in the literatures \cite{22,23,24}.

The ``$L$--$\sigma$" relation between the luminosity $L($H$\beta)$
in H$\beta$ of a source and its ionized gas velocity dispersion can
be expressed as \cite{19}
\begin{equation}
\log L ( \textrm{H}\beta) = \alpha \log \sigma(\textrm{H}\beta)+\kappa,
\end{equation}
where $\alpha$ is the slope and $\kappa$ is the intercept. The
$L(\textrm{H}\beta)$ is obtained from the reddening corrected flux
density $F(\textrm{H}\beta)$ which only bases on a general equation
$L(\textrm{H}\beta) = 4\pi D^{2}_L F(\textrm{H}\beta)$. Thus, the
equation above can be written as a relation of the observed flux
density
\begin{equation}
\log D_{L,HII}(z) = 0.5[\alpha \log \sigma(\textrm{H}\beta)- \log F ( \textrm{H}\beta)+\kappa]-25.04.
\end{equation}
Analogous to SN Ia applied in cosmology, the $\alpha$ and $\kappa$
parameters should also be optimized with the assumed cosmological
model parameters. Fortunately, Wu \textit{et al.} used the
measurements of Hubble parameters from cosmic clocks
(model-independent) to calibrate $\alpha$ and $\kappa$, and
demonstrated that the calibrated values $\alpha=5.12\pm 0.08$ and
$\kappa=33.08\pm 0.13$ are reliable for cosmological applications
\cite{24}. In this work, we will adopt these values with their
corresponding uncertainties to get the luminosity distance.

The catalog of spectral and astrometric data from HII galaxies and
extragalactic HII regions contain more than 100 sources by far, and
its statistical properties can be preliminarily considered in
cosmology. In this work, we will use the current observations of 156
HII objects compiled by Terlevich \textit{et al.} \cite{19} which
contain 25 high redshift HII galaxies sources, 107 local HII
galaxies sources, and 24 extragalactic HII regions sources covering
redshift range $0<z<2.33$. This dataset is larger than the source
samples used by Plionis \textit{et al.} \cite{22} and is more
complete than the high redshift data used by Melnick \textit{et al.}
\cite{21}. Full information (including name of the source, redshift,
flux density, and turbulent velocity with corresponding
observational  uncertainties) about the sample of 156 HII regions
can be found in Table 1 of the work \cite{20}.

\subsection{Angular diameter distances measured from compact structure of radio quasars}

Quasars, among the most distant objects in the universe, have great
potential as distance indicators. From an observational point of
view, there are currently two types of quasar data that can be served as
cosmological probes, i.e., the non-linear relation between the
ultraviolet and X-ray fluxes of the quasar to construct the Hubble
diagram \cite{13,14,15}, and the angular size-distance relation of
ultra-compact structures object in radio quasars (QSO) as the
standard ruler of cosmology from the very-long-baseline
interferometry (VLBI) observations \cite{16}. The first type of
quasar data provides the luminosity distance, but not the
angular diameter distance, directly. Moreover, although the sample collected
by \cite{13} contained 1598 suitable quasars and redshift reaches to
$z\sim5.5$, the sample itself exhibits a large intrinsic dispersion.
Take these factors into consideration, we will use the radio quasar
sample to obtain the angular diameter distance information.

The angular size-distance relation in compact radio quasar for
cosmological inference was first proposed by Kellermann \textit{et
al.} \cite{25}, in which he tried to obtain the deceleration
parameter with 79 compact radio sources from VLBI at 5 GHz.
Whereafter, Gurvits \cite{26} extended this method and attempted to
investigate the dependence of characteristic size on luminosity and
redshift based on 337 Active Galactic Nucleuses (AGNs) observed at
2.29 GHz
 \cite{27}. In the subsequent analysis, the literature
\cite{26} adopted the modulus of visibility $\Gamma=S_c/S_t$  to
redefine angular size of radio sources $\theta$, which can be
expressed by $\theta(z)=2\sqrt{-\ln\Gamma \ln 2}/\pi B_{\theta}$,
where $B_{\theta}$ is interferometer baseline measured in
wavelengths, $S_c$ and $S_t$ are correlated flux density and total
flux density, respectively. Based on a simple relation between the
angle and distance with the intrinsic linear size $l_m$ of the
compact structure in radio quasars, the angular size $\theta(z)$ can
be written as
\begin{equation}
D_{A,QSO}(z)=\frac{l_m}{\theta(z)},
\end{equation}
where $l_m=lL^{\beta}(1+z)^n$ describes the apparent distribution of
radio brightness within the core, $l$ is the linear size scaling
factor, $L$ is the intrinsic luminosity of the source, and $\beta$
and $n$ represent the possible dependence of the intrinsic linear
size of the source on luminosity and redshift, respectively. With
the gradually refined selection technique and observations, as well
as the elimination of systematic errors caused by various aspects,
Cao \textit{et al.} \cite{16} compiled milliarcsecond compact radio
sample of 120 intermediate-luminosity $(10^{27}W/Hz < L < 10^{28}
W/Hz)$ quasars with reliable measurements of the angular size of the
compact structure covering the redshift range $0.46<z<2.76$ from
VLBI survey at 2.29 GHz. They showed that $l_m$ is independent of
redshift and luminosity ($|\beta|\approx10^{-4},
|n|\approx10^{-3}$), which suggests that it can be used to
cosmological studies. However, the current problem is how to
determine the value of $l_m$. In the subsequent analysis, the linear
size, without pre-assuming a cosmological model, was determined to
be $l_m=11.03\pm0.25$ pc by Cao \textit{et al.} based on the
$D_A(z)$ reconstruction from $H(z)$ data obtained from cosmic
chronometers\cite{16}. The calibrated intrinsic length and
cosmological application of this sample had obtained stringent
constraints on both the matter density parameter $\Omega_m$ and the
Hubble constant $H_0$, which are consist with Planck 2018
observation \cite{3}. The ultra-compact structures in radio quasars
for exploring other cosmological models has been investigated in
many literatures \cite{28,29,30,Xu18,31,32}.Therefore, it is
reasonable to ask whether the derived angular size depends on the
intrinsic luminosity of the radio quasar, and consequently affecting
testing the validity of CDDR. In fact, the derived angular size is
obtained by a ratio of correlated and total flux densities, i.e.,
the modulus of visibility $\Gamma=S_c/S_t$. Therefore, from the
perspective of observation, the intrinsic luminosity of the radio
quasar does not affect the effectiveness of CDDR testing.

\subsection{Reconstructions based on Gaussian Process and Artificial Neural Network}

From a theoretical perspective, one can directly achieve CDDR
testing by combining the $L$--$\sigma$ relation in HII regions with
the angular size-distance relation of compact radio sources.  From the
observational point of view, however, there is currently a lack of
data samples. Not only of the HII region, but also samples of quasars.
Although their redshifts cover each other well, there are very
few of them meeting the same redshift at the same time. In order to
achieve the CDDR testing and obtain convincing results, we consider
two non-parameterized technologies, Gaussian Process (GP) and
Artificial Neural Network (ANN), to reconstruct the HII galaxy
Hubble diagram and ultra-compact structure of radio quasar sources
data, respectively. There is no reason to favor one technology over
another, but mutually consistent results for different parameterized
technique would strengthen the robustness of the conclusion.

\textit{Gaussian Process.---} The Gaussian Process (GP) is a random
process defined in the continuous domain, which can be regarded as a
set of all random variables in the continuous domain, and any single
or multiple random variables satisfy one-dimensional Gaussian
distribution or multi-dimensional Gaussian distribution \cite{33}.
The GP can be determined by a mean function $m(\textbf{x})$ and a
covariance function $k(\textbf{x},\textbf{x}_{\ast})$ (also called
kernel function). The GP defines a priori function, one can assume
that for a given $\textbf{x}$, $\textbf{y}$ there follows a
distribution
$p(\textbf{y}|\textbf{x})=\mathcal{N}(\textbf{y}|m,\textbf{K})$,
which $\textbf{K}=k(\textbf{x},\textbf{x}_{\ast})$. The purpose of
GP is to learn a mapping function $f$ from $\textbf{x}$  to
$\textbf{y}$ through $\textbf{x}$, $\textbf{y}$. Then, for the given
new $\textbf{x}_{\ast}$, one can predict
$\textbf{y}_{\ast}=f(\textbf{x}_{\ast})$. According to priori
distribution of GP, the joint distribution
$p(\textbf{y},\textbf{y}_{\ast}|\textbf{x},\textbf{x}_{\ast})$ of
observed data $\textbf{y}$ and forecast data $\textbf{y}_{\ast}$ are
given by \cite{34}
\begin{equation}\label{initialcov}
\left(
\begin{array}{c}  $\textbf{y}$  \\  \textbf{y}_{\ast}
\end{array}\right)\sim \mathcal{N}
\left(
\left(
\begin{array}{c} m \\ m_{\ast}
\end{array}\right),\left(
\begin{array}{cc} \textbf{K}_{\textbf{y}} &\textbf{K}_{\ast} \\ \textbf{K}_{\ast}^T &  \textbf{K}_{\ast\ast}
\end{array}\right)
\right),
\end{equation}
where $\textbf{K}_{\textbf{y}}=k(\textbf{x},\textbf{x})$,
$\textbf{K}_{\ast}=k(\textbf{x},\textbf{x}_{\ast})$ and $\textbf{K}_{\ast\ast}=k(\textbf{x}_{\ast},\textbf{x}_{\ast})$.
According to the joint distribution of $\textbf{y}$ and $y_{\ast}$, one can get the
conditional distribution
\begin{equation}
p(\textbf{y}_{\ast}|\textbf{x}_{\ast},\textbf{x},\textbf{y})=\mathcal{N}(\textbf{y}_{\ast}|
\mu_{\ast},\Sigma_{\ast}),
\end{equation}
where $\mu_{\ast}=\textbf{K}_{\ast}^T\textbf{K}_{\textbf{y}}^{-1}\textbf{y}$ and  $\Sigma_{\ast}=\textbf{K}_{\ast\ast}-\textbf{K}_{\ast}^T\textbf{K}_{\textbf{y}}^{-1}\textbf{K}_{\ast}$
are the expectation vectors and covariance matrices of the posterior prediction distribution, respectively.

The kernel function has many choices, such as square exponential function. We take the Mat\'{e}rn ($\nu = 9/2$) covariance function here, because it can provide more reliable results when using GP to reconstruct function \cite{35}
\begin{align}
k(z,\tilde z) = &~{\sigma _f}^2\exp\left( - \frac{{3\left| {z - \tilde z} \right|}}{\ell }\right) \nonumber \\
      &~~\times\Big[1 + \frac{{3\left| {z - \tilde z} \right|}}{\ell } + \frac{{27{{(z - \tilde z)}^2}}}{{7{\ell ^2}}} \nonumber \\
     &~~ + \frac{{18{{\left| {z - \tilde z} \right|}^3}}}{{7{\ell ^3}}} + \frac{{27{{(z - \tilde z)}^4}}}{{35{\ell ^4}}}\Big],
\end{align}
where $\ell$ denotes the characteristic length scale in $x$-direction and
$\sigma_f$ is the signal variance in $y$-direction. One can see
that, except for the hyper parameters in kernel function, there is no
parameter estimation that was involved in the final prediction of
the $\textbf{y}_{\ast}=f(\textbf{x}_{\ast})$. It should be emphasized here that whenever one performs the GP regression, the hyperparameters should be optimized by GP with the observed data set, along with other parameters of interest (cosmological parameters or not). Therefore, the proper way of performing such GP analysis is to treat the GP hyperparameters on the same footing as the cosmological parameters, i.e. varying all relevant parameters together and sampling the joint posterior. Such procedure, which guarantees that the
reconstructed function is independent of the initial hyperparameter
settings, has been extensively applied in different cosmological studies \cite{36,37,38,39}, especially
high-fidelity constraints on the spatial curvature parameter and Hubble constant \cite{Colgain21,Dhawan21}.

In this analysis, we use Gaussian Processes in Python (GaPP)
\footnote{http://www.acgc.uct.ac.za/seikel/GAPP/index.html} to realize the
reconstruction of different functions. For the
HII regions sample, the reconstructed logarithmic luminosity
distance $\log D_{L,HII}(z)$, as a function of logarithmic redshift
$\log z$, with the estimation of their $1\sigma$ confidence regions
are shown in the top panel of Fig.~1. Meanwhile, we also reconstruct
function $\theta(z)$ from compact radio sources observations, and
final results with their corresponding $1\sigma$ uncertainties are
shown in the bottom panel of Fig.~1. We reconstruct 1000 points for
HII regions and compact radio sources, respectively. From
this figure, one can see that the $1\sigma$ uncertainty from GP
reconstruction is smaller than that of individual data points. Such
issue has been extensively discussed in the recent works
\cite{33a,33b}. More specifically, the final reconstructed
confidence region depends on three factors, i.e., the observed
errors of data, the optimization of hyper parameters of the GP
method, and the product of the covariance matrixes
$\mathbf{K}_{\ast}\mathbf{K_y}^{-1}\mathbf{K}_{\ast}^{T}$ between
the predicted and current observed points. It should be noted that,
if
$\mathbf{K}_{\ast}\mathbf{K_y}^{-1}\mathbf{K}_{\ast}^{T}>\sigma_f$,
the uncertainty of the predicted point will be less than
uncertainties of observed points when there is a large correlation
between the data. One can clearly see from Eq.~(6) that the
correlation between $z$ and $\tilde z$ will be large when $z-\tilde
z$ less than $\ell$. Such condition, which is satisfied by most of
the HII galaxy and quasar data points in our study, will result in
smaller 1$\sigma$ confidence region from GP. We refer the reader to
Ref.~\cite{33} for further details on this issue.

\begin{figure}
\begin{center}
\includegraphics[width=0.9\linewidth]{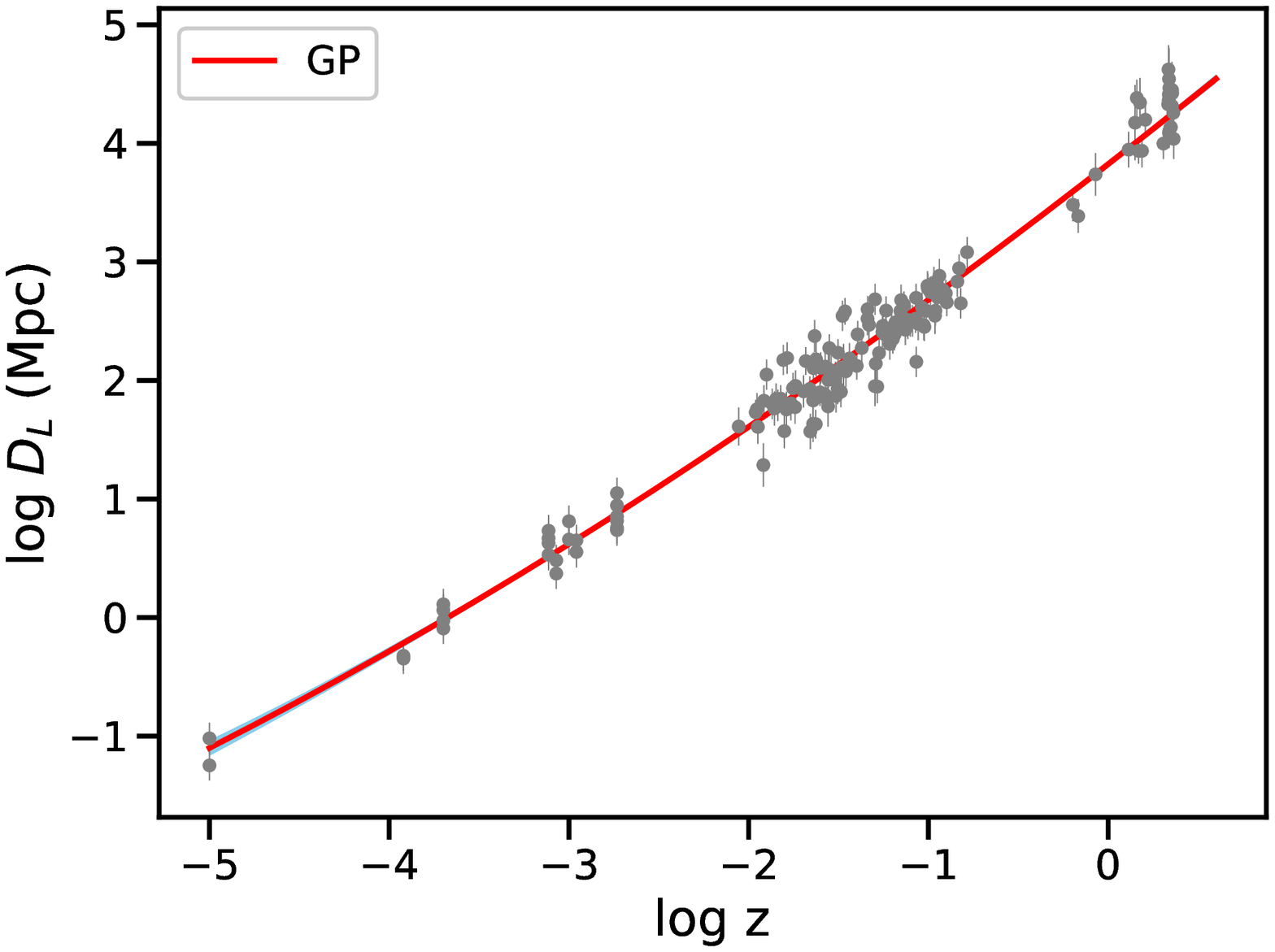}\\
\vskip 0.2in
\includegraphics[width=0.9\linewidth]{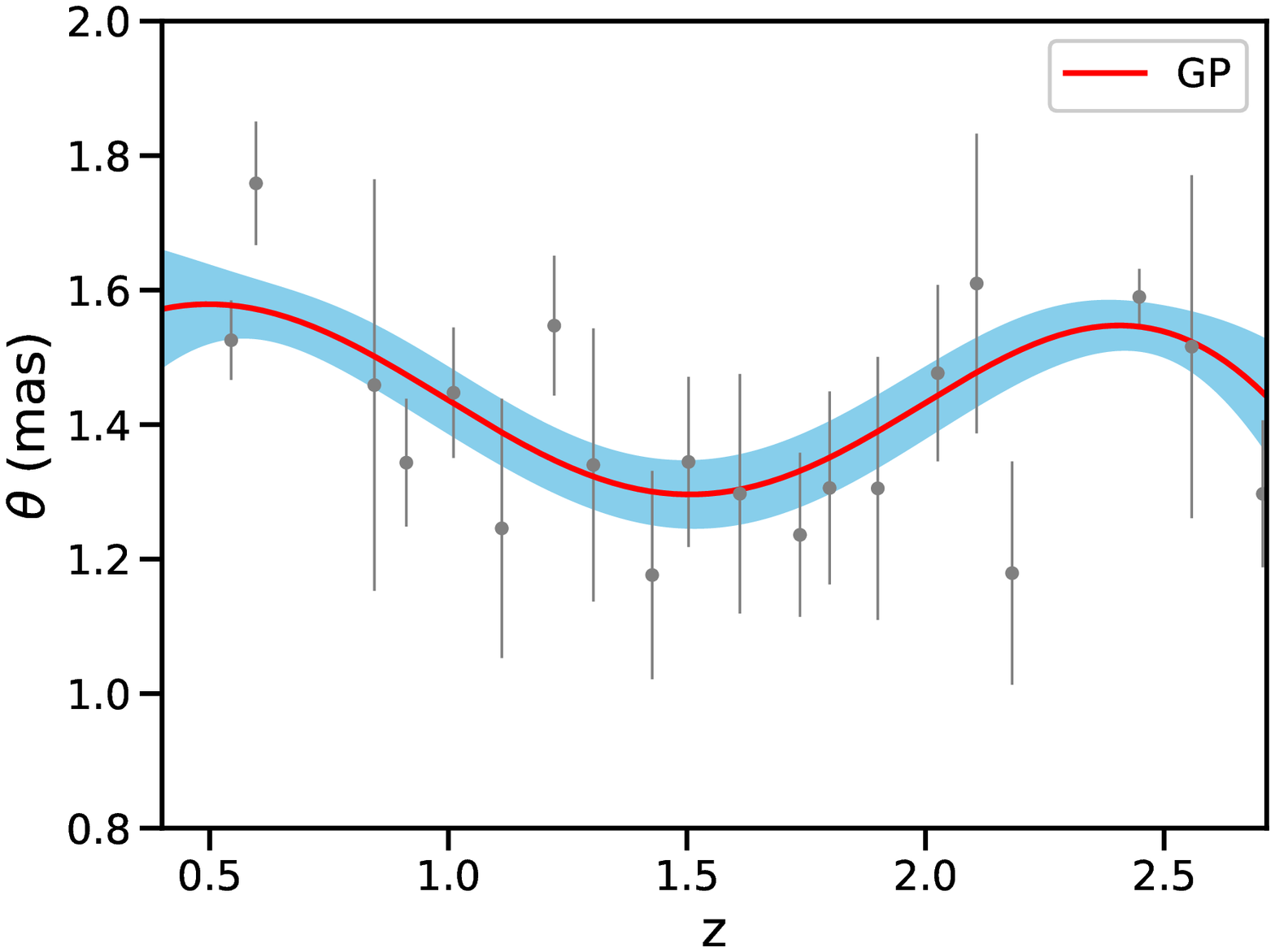}
\end{center}
\caption{{\it Top panel:} The reconstructed function $\log
D_{L,HII}(z)$ with corresponding $1\sigma$ errors by using GP (red
line), and the gray dots dots with error bars represent measurements
of HII regions;  {\it Bottom panel:} The GP-reconstructed function
$\theta(z)$ from compact radio sources observations.}
\end{figure}

\textit{Artificial Neural Network.---} For the second nonparametric
approach, we turn to Artificial Neural Network (ANN) and show the
reconstructed function from observational data. With the development
of computer hardware in the recent ten years, machine learning
technology has been gradually applied to many research fields in
astronomy, and shown excellent potential for solving cosmological
problems, such as analyzing gravitational waves \cite{40,41} and
constraining cosmological parameters \cite{42,43,44,45,46,47}.

The main purpose of an ANN is to construct an approximate function
or map that correlates the input data with the output data. The ANN has been
shown to be "universal approximator" that can represent a wide
variety of functions \cite{48,49}.  The ANN is made up of neurons,
which are very simple elements that receive digital input. Generally
speaking, the artificial neural network consists of an input layer,
one or more hidden layers and an output layer. Each layer takes a
vector from the previous layer as input, applies a linear
transformation and a nonlinear activation function to the input, and
propagates the current result to the next layer. Formally, in a
vectorized way \cite{50}
\begin{equation}
\mathbf{z_{i+1}}=\mathbf{x_i}W_{i+1}+\mathbf{b_{i+1}},
\end{equation}
\begin{equation}
\mathbf{x_i}=f(\mathbf{z_{i+1}}),
\end{equation}
where $\mathbf{x_i}$ is the input vector at the $i$th layer,
$W_{i+1}$ and $\mathbf{b_{i+1}}$ are linear weights matrix and the
offset vector which need to be optimized, $\mathbf{z_{i+1}}$ is the
output vector after linear transformation, and $f$ is the activation
function. Here, the Exponential Linear Unit (ELU) is acted as the
activation function \cite{51}, which is given by
\begin{align}\label{eq:elu}
&f(x) =
\begin{cases}
x & x > 0 \\
\alpha (\exp(x)-1) & x \leq 0
\end{cases} \quad ,
\end{align}
where $\alpha$ denotes the hyper-parameter that controls the value to
which an ELU saturates for negative net inputs. Compared to other
activation functions (such as the rectified linear and the leaky
rectified linear), when the network exceeds five layers, ELU can not
only improve the learning speed, but also have better generalization
performance \cite{51}.

The ANN equals to a function $f_{W,\mathbf{b}}(\mathbf{x})$. The goal
of ANN is to make its output to be as close as possible to the
target value $\mathbf{y}$. Then, according to the difference between
the predicted value $f_{W,\mathbf{b}}(\mathbf{x})$ of the current
network and the target value $\mathbf{y}$, the weight matrix of each
layer needs to be constantly updated for minimize the difference,
which is defined by a loss function $\mathcal{L}$. The method used
is gradient descent, that is, by constantly moving the loss value to
the opposite direction of the current corresponding gradient to
reduce the loss value. Formally, in a vectorized way \cite{52}
\begin{eqnarray} \nonumber
\frac{\partial \mathcal L}{\partial x_{i + 1}}&=f^{\prime}(x_{i + 1})\frac{\partial \mathcal L}{\partial x_{i+1}},\\ \nonumber
\frac{\partial \mathcal L}{\partial W_{i + 1}}&=x_{i}^T\frac{\partial \mathcal L}{\partial z_{i+1}},\\ \nonumber
\frac{\partial \mathcal L}{\partial x_{i}}&= W_{i+1}^T \frac{\partial \mathcal L}{\partial z_{i + 1}}, \\
\frac{\partial \mathcal L}{\partial {b}_{i+1}} &= \left(\frac{\partial \mathcal L}{\partial z_{i+1}}\right),
\end{eqnarray}
where the operator $\partial$ denotes partial derivatives, and
$f^\prime$ is the derivative for the nonlinear function $f$.

According to the publicly released code by the work \cite{50}, which
explicitly describe the ANN method, we use the module called
Reconstruct Functions with ANN (ReFANN)
\footnote{https://github.com/Guo-Jian-Wang/refann} to perform the
reconstruction of the HII regions and radio quasars data-sets.
Similarly, we show the reconstructed function $\log D_{L,HII}(z)$ by
using ANN method as a function of logarithmic redshift $\log z$ with
the estimation of $1\sigma$ confidence region in the top panel of
Fig.~2. For compact radio sources, the reconstructed function
$\theta(z)$ with corresponding $1\sigma$ uncertainties by using ANN
is given in the bottom panel of Fig.~2. Similarly, we also
reconstruct 1000 points for HII regions and compact radio sources,
respectively. Compared to GP technology, the ANN method does not
assume random variables that satisfy the Gaussian distribution,
which is a completely data driven approach. It is
interestingly to note that the uncertainties of the data
reconstructed by ANN are almost equal to that of the observations.
Therefore, the $1\sigma$ confidence region reconstructed by ANN can
be considered as the average level of observational error. We refer
the reader to Ref.~\cite{46} for further details on this issue.

\begin{figure}
\begin{center}
\includegraphics[width=0.9\linewidth]{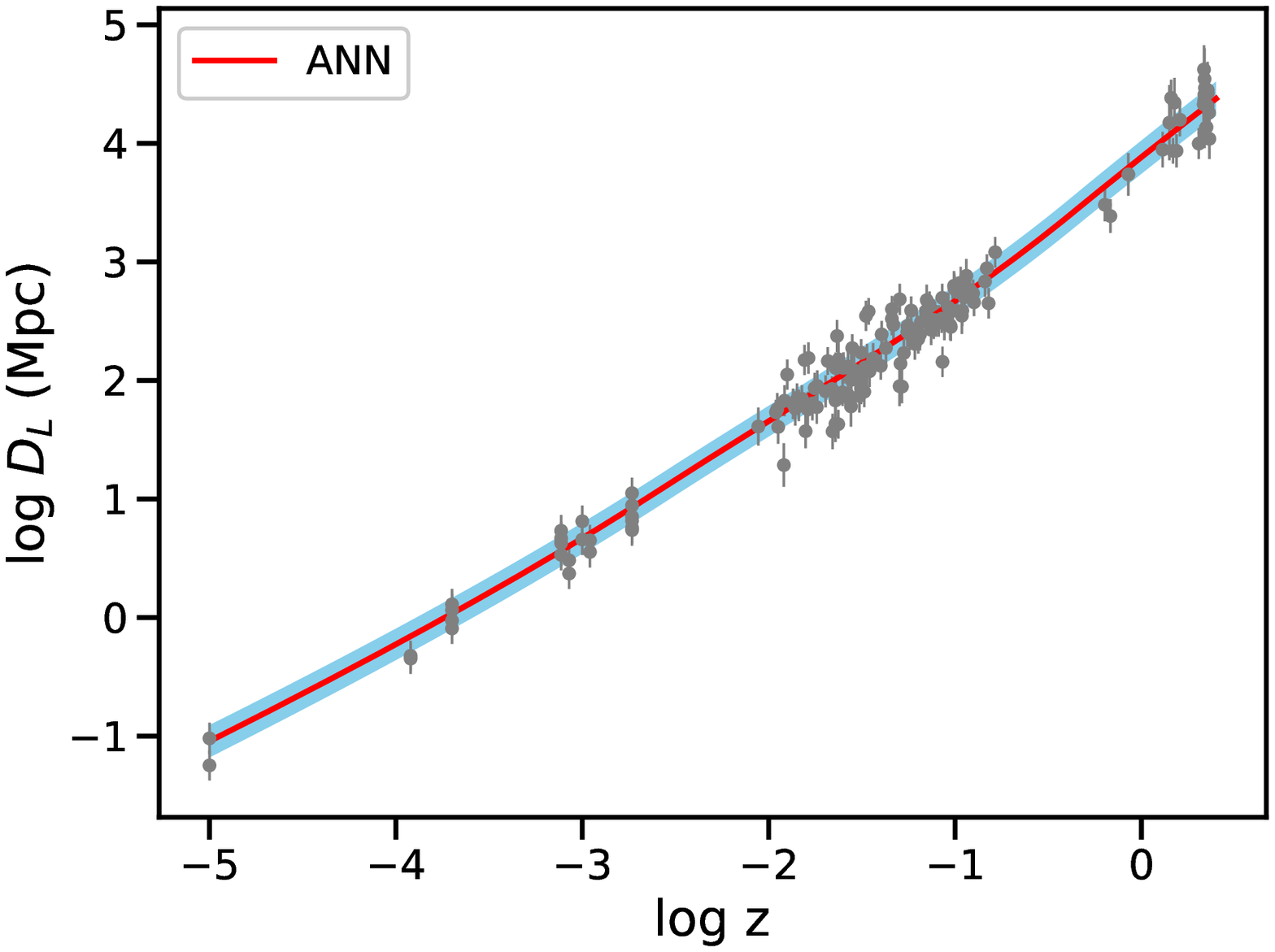}\\
\vskip 0.2in
\includegraphics[width=0.9\linewidth]{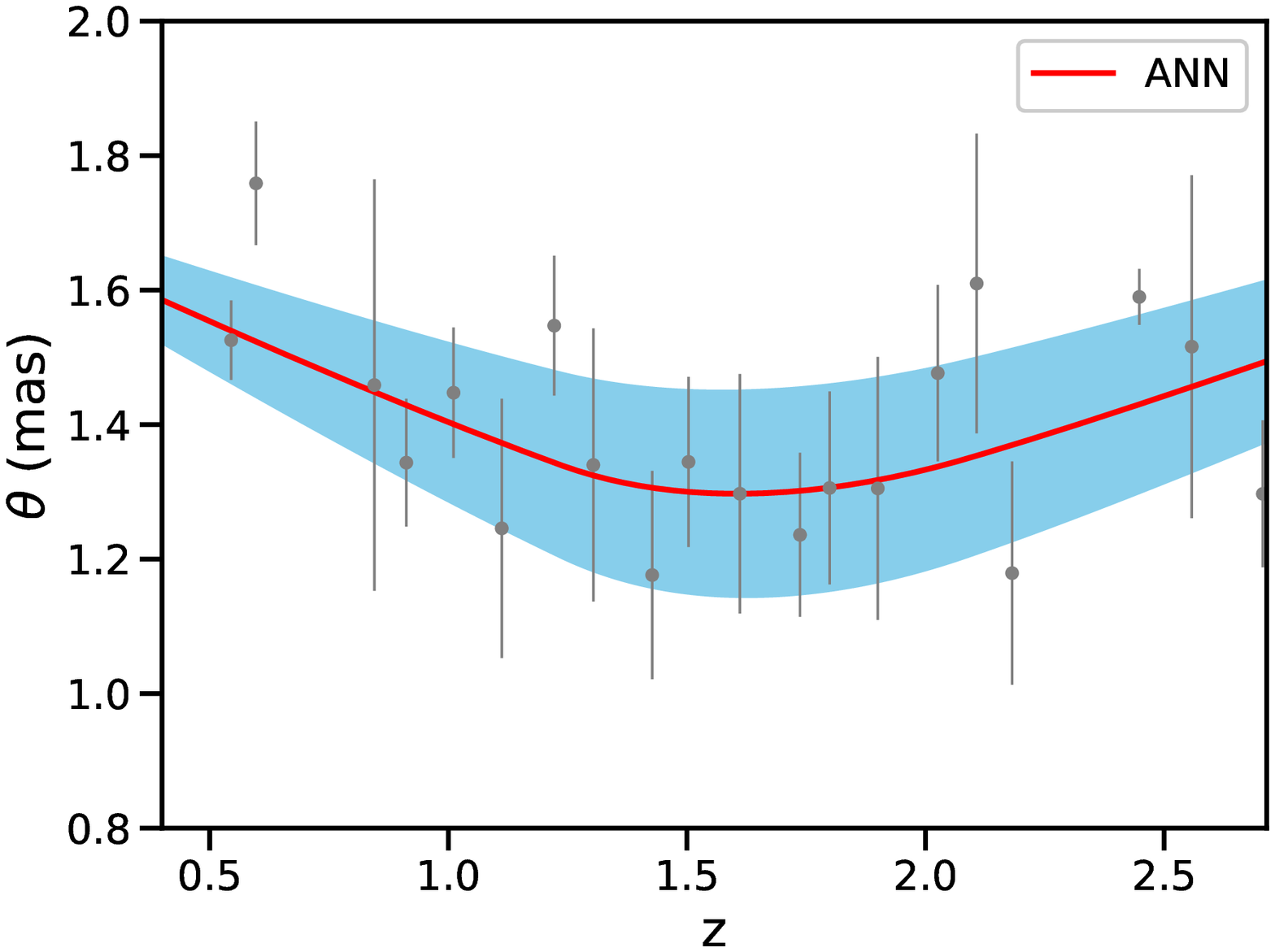}
\end{center}
\caption{{\it Top panel:} The reconstructed function $\log
D_{L,HII}(z)$ with $1\sigma$ errors by using ANN (red line), and the
gray dots dots with error bars represent measurements of HII
regions;  {\it Bottom panel:} The reconstructed  function $\theta(z)$
by using ANN from compact radio sources observations.}
\end{figure}


\subsection{Testing the validity of CDDR} \label{subsec:CDDR_Test}

In order to test the validity of CDDR, we use the
reconstructed data to directly perform a model-independent test of
CDDR, i.e., we do not adopt any parameterized form to quantify the
CDDR, which is given by the following form \cite{53,54}
\begin{equation}\label{eq:DDR}
\eta(z)=\frac{D_L(z)}{D_A(z)(1+z)^2}.
\end{equation}
Note that any statistically significant deviation from $\eta(z)=1$
could indicate possible violation of the three basic CDDR
assumptions. Furthermore, we turn to four parameterized forms of
CDDR, which have been extensively discussed in the quoted papers
\cite{Cao11b}
\begin{equation}\label{eq:DDReta}
\eta_{th}(z,\eta_j)=\left\{\begin{array}{l@{\hspace{1cm}}l}
1+\eta_{0}, \\
1+{\eta_1}z, \\
1+\eta_{2}z/(1+z), \\
1+\eta_{3}z+\eta_{4}z^2.\\
\end{array}\right.
\end{equation}
Note that the first parameterized form is independent of redshift,
therefore we can compare it to other parameterized forms to check
its possible dependency on the redshift. In general, $\eta(z)$ can
be treated as parameterized functions of the redshift. In this work,
we also use other general parametric representations for a possible
redshift dependence of CDDR including two one-parameter expressions
and a two-parameter parametrization. The consistency of the results
under different parameterized forms will enhance the robustness of
the conclusion.

From the observational perspective, we obtain the luminosity distance
$D_L(z)$ through the "$L$--$\sigma$" of HII galaxies and
extragalactic HII regions, and the diameter distance $D_A(z)$ can be
derived from compact structure in radio quasars. For a given
$D_{L,HII}$ data point, the angular diameter distance $D_{A,QSO}$
should be observed at the same redshift. To avoid introducing
additional systematic errors, a cosmological model-independent
selection criterion is considered. We take $|z_{HII}-z_{QSO}|<0.005$ in our
analysis \cite{6,7}. If one only consider the actual observational
sample, it is difficult to achieve a rigorous CDDR test and get
convincing results. That is the reason why we used two
non-parameterized techniques mentioned above to reconstruct the
data. Using the reconstructed samples, we are able to have a
one-to-one matching between HII regions and compact structure in
radio quasars . After executing the redshift selection criterion by
using reconstructed data samples, 892 data are remained.
Subsequently, the observed $\eta_{obs}(z)$ can be represented by
following form
\begin{eqnarray}
\eta_{obs}(z)&=&\frac{D_{L,HII}}{D_{A,QSO}(1+z)^{2}}\\
&=&\frac{\theta}{l_m(1+z)^{2}}10^{0.5[\alpha \log \sigma(\textrm{H}\beta)- \log F( \textrm{H}\beta)+\kappa]-25.04}. \nonumber
\end{eqnarray}
Uncertainties have been assessed from the standard uncertainty
propagation formula, based on (uncorrelated) uncertainties of
observable quantities. The total uncertainty budget includes the
ionized gas velocity dispersion $\sigma_{\log
\sigma(\textrm{H}\beta)}$, flux density $\sigma_{\log
F(\textrm{H}\beta)}$ and additional systematic errors introduced from
the calibrations of $\alpha$ and $\kappa$ in HII regions data,
the angular size $\sigma_{\theta}$, and additional systematic errors
introduced in the calibrations of linear size $l_m$ in radio
quasars. So the total uncertainty of $\eta_{obs}(z)$ can be
expressed as
\begin{equation}\label{eq:etaobssig}
\sigma_{\eta_{obs}} =\sqrt{\sigma_{\log
\sigma(\textrm{H}\beta)}^2+\sigma_{\log
F(\textrm{H}\beta)}^2+\sigma_{\alpha}^2+\sigma_{\kappa}^2+\sigma_{\theta}^2+\sigma_{l_m}^2}.
\end{equation}

In order to determine the best fitting CDDR parameters and
corresponding uncertainties, we use the Bayesian statistical methods
to obtain the posterior probability density function of the CDDR
parameters $\eta_j$ ($j = 0,.. ,4$) corresponding to four
parametrization forms of Eq.~(\ref{eq:DDReta}). The posterior probability
density function is given by
\begin{equation}
p(\mathbf{\eta}|obs\,\ data)\propto \mathcal{L}(\mathbf{\eta}, obs\,\ data )\times p(\mathbf{\eta}),
\end{equation}
where $\mathcal{L}$ is the likelihood function, and has a following form of
\begin{equation}
\mathcal{L}=\prod_{i}^{i=892}\frac{1}{\sqrt{2\pi}\sigma_\eta}{\rm
exp}\left[-\frac{1}{2}\frac{(\eta_{obs}(z)-\eta_{th}(z,\eta_j))^2}{\sigma_\eta^2}\right],
\end{equation}
and $p(\eta)$ is the prior, and assumed the following uniform
distribution: $p(\eta_j)=U[-1,1]$. We use the Python module $emcee$
\cite{55} to perform the Markove Chain Monte Carlo (MCMC) analysis.

%
\begin{figure}
\begin{center}
\includegraphics[width=0.9\linewidth]{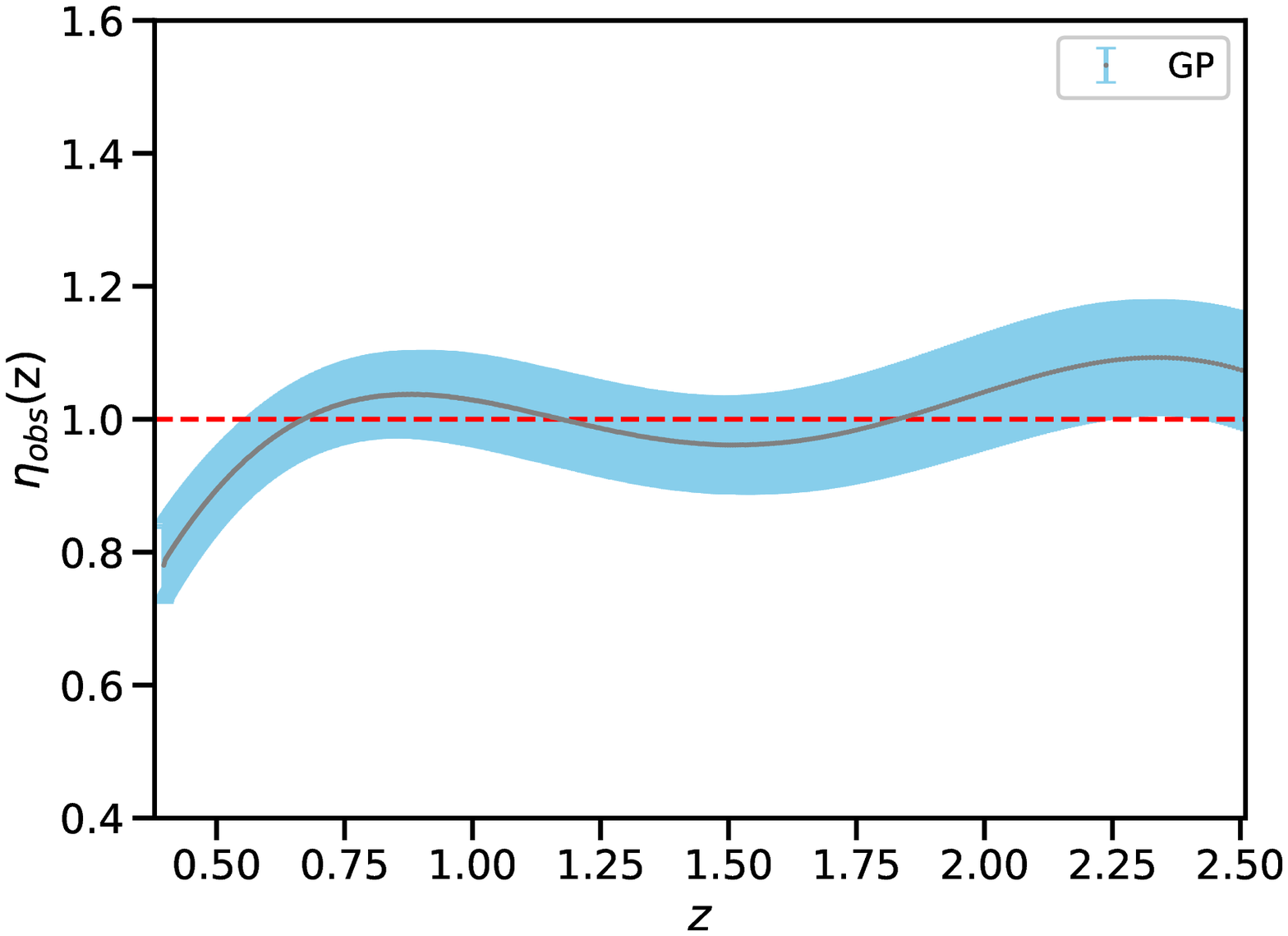}
\end{center}
\caption{Cosmic distance duality relation $\eta(z)$ for the
GP reconstruction from the HII galaxy and compact radio quasar
sample. The dashed line at unity is $\eta(z)=1$, the solid line is
the ANN fit, and the shaded region is the $1\sigma$ GP errors.}
\end{figure}

\begin{figure}
\begin{center}
\includegraphics[width=0.9\linewidth]{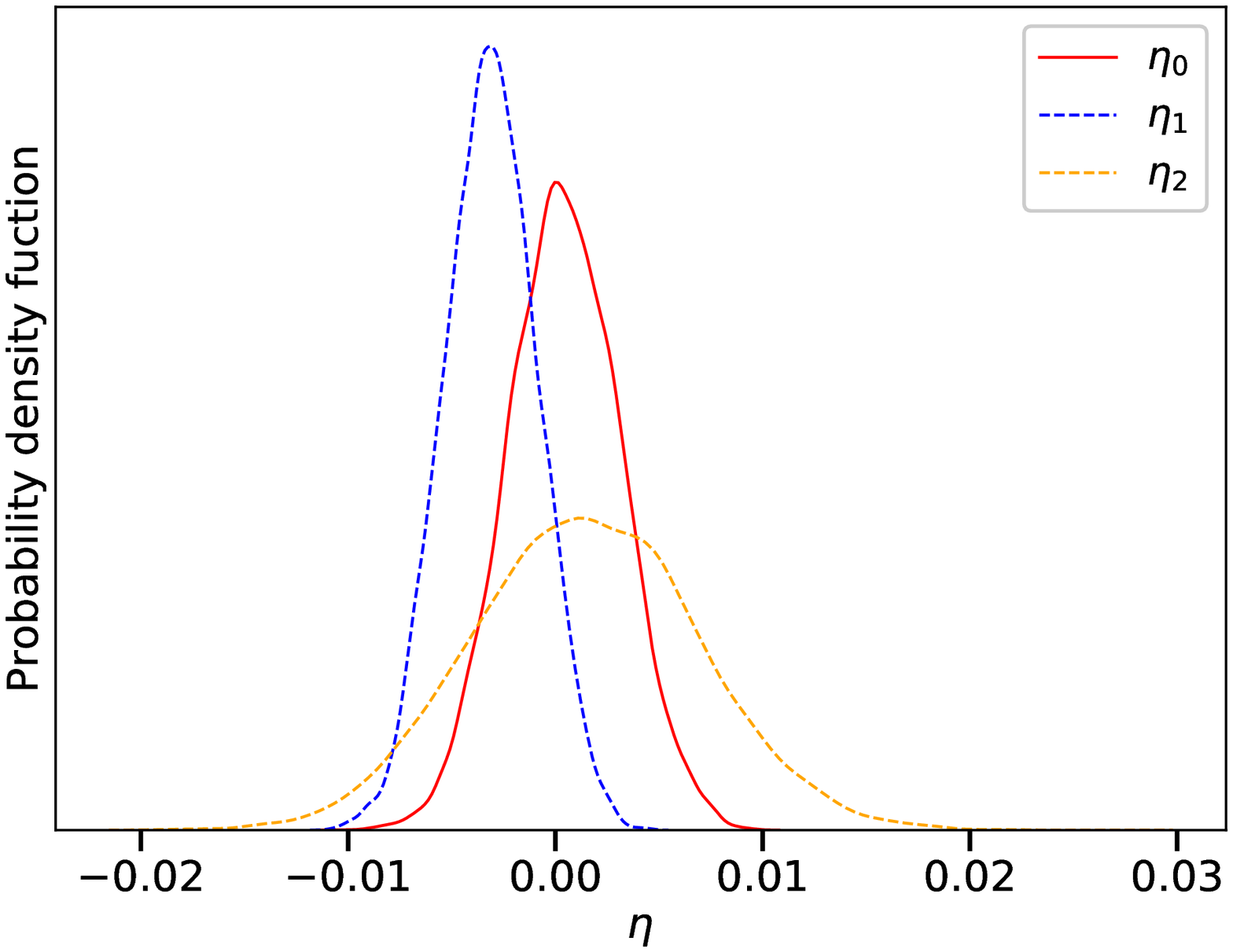}
\end{center}
\caption{The posterior probability density function with three
parameterized forms $\eta_j$ ($j = 0, 1 ,2$)  by using GP
reconstructed HII galaxy and compact radio quasar samples.}
\end{figure}

\begin{figure}
\begin{center}
\includegraphics[width=0.8\linewidth]{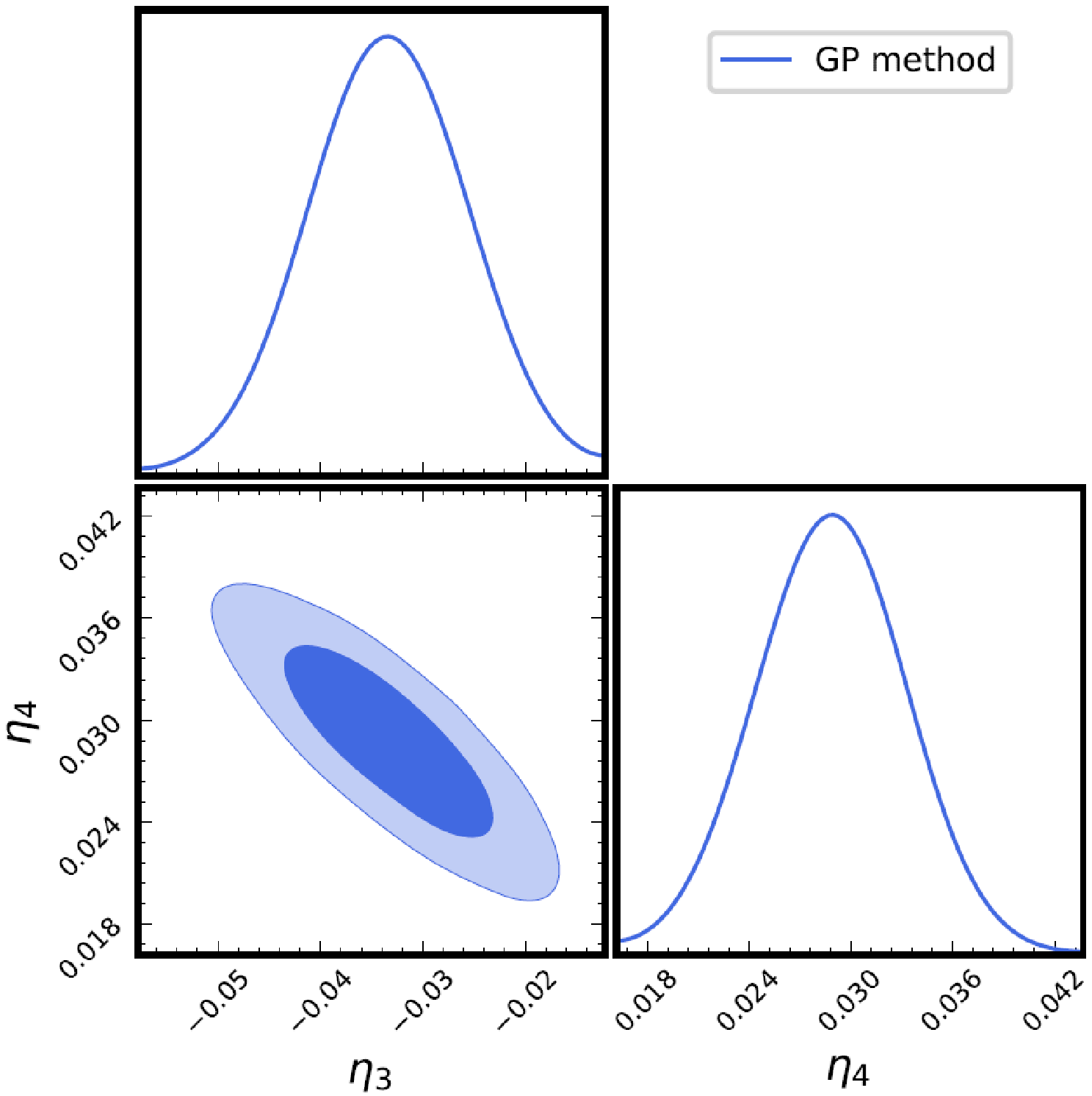}
\end{center}
\caption{ The 1D and 2D marginalized probability distributions for
the fourth CDDR parameters $\eta_3$ and $\eta_4$ by using GP
reconstructed HII galaxy and compact radio quasar samples.}
\end{figure}

\section{Results and Discussion}\label{sec:result}

\begin{table*}
\begin{center}
\begin{tabular}{c| c c c c c}
\hline\hline
$\eta_{j}$(z)+GP method  & $\eta_{0}$ & $\eta_1$ &$\eta_{2}$ &$\eta_{3}$ &$\eta_{4}$  \\
\hline
  $1+\eta_{0}$ & $0.001\pm0.003$ & $\Box$ & $\Box$ & $\Box$ & $\Box$ \\
\hline
$1+{\eta_1}z$  & $\Box$ & $-0.003\pm0.003$  &$\Box$ & $\Box$  & $\Box$ \\
\hline
$1+\eta_{2}z/(1+z)$   & $\Box$ & $\Box$  & $ 0.001\pm0.005$  &$\Box$ &$\Box$ \\
\hline
$1+\eta_{3}z+\eta_4z^2$   & $\Box$ & $\Box$  & $\Box$ &$-0.033\pm0.006$ &$0.029\pm0.003$ \\
\hline \hline
$\eta_{j}$(z)+ANN method  & $\eta_{0}$ & $\eta_1$ &$\eta_{2}$ &$\eta_{3}$ &$\eta_{4}$  \\
\hline
  $1+\eta_{0}$ & $0.031\pm0.016$ & $\Box$ & $\Box$ & $\Box$ & $\Box$ \\
\hline
$1+{\eta_1}z$  & $\Box$ & $0.028\pm0.014$  &$\Box$ & $\Box$  & $\Box$ \\
\hline
$1+\eta_{2}z/(1+z)$   & $\Box$ & $\Box$  & $0.031\pm0.033$  &$\Box$ &$\Box$ \\
\hline
$1+\eta_{3}z+\eta_4z^2$   & $\Box$ & $\Box$  & $\Box$ &$0.033\pm0.043$ &$0.027\pm0.025$ \\
\hline
\hline
\end{tabular}
\caption{Constraints on the CDDR parameters for four types of
parameterized forms, in the framework of GP and ANN technologies.}
\label{SIE_table}
\end{center}
\end{table*}

Let's start with the reconstruction of HII galaxy and radio quasar
sample using GP. In Fig.~3 we show a particular realization
of the reconstructed CDDR, along with the case of $\eta(z)=1$
(dashed green line) and the corresponding best-fit (solid colored
line) for the GP. Our results indicate that there is no obvious
deviation from $\eta(z)=1$ at $1\sigma$ confidence level. In the
higher redshift region, due to the lack of observational data the
errors of reconstruction become larger and the statistical
significance of CDDR reconstruction is affected. Such finding is
well consistent with that obtained in the recent works
\cite{12a,12b,12c,12d}. Focusing on the different parameterized
form of $\eta_{obs}(z)$, the numerical results by using GP
reconstructed technology for three parameterized forms $\eta_j$ ($j
= 0, 1 ,2$) are summarized in Table 1, and the posterior probability
density functions are shown in Fig.~4. For the first parameterized
form, the best fitting value with 1$\sigma$ error is
$\eta_0=0.001\pm0.0031$, which demonstrates that there is no
evidence for the dependence relation between the CDDR parameter and
redshift. Working on the second parameterized form, we obtain
$\eta_1=-0.003\pm0.003$, which contains zero value within $\sim
1.2\sigma$ confidence level. Considering the redshift coverage of
our CDDR test ($z\sim 2.3$), the third parametrization may
effectively avoid the possible divergence at high redshift. In this
case, the best fitting value $\eta_2$ with 1$\sigma$ confidence
level is $\eta_2=0.001 \pm0.005$. Meanwhile, for the two-parameter
form, the graphic representation and numerical results of the
constraints on the CDDR parameters ($\eta_3$, $\eta_4$) are shown in
Fig.~5 and Table 1. One can clearly see that the CDDR seems to be
violated at 1$\sigma$ confidence level, $\eta_3=-0.033\pm0.006$ and
$\eta_4=0.029\pm 0.003$. However, one should be noted that the
degeneracy between $\eta_3$ and $\eta_4$ is strong, and they are in
negative correlation. If $\eta_3$ increases to zero, then $\eta_4$
will go back to zero, which means that the strong degeneracy between
them affects our test of CDDR validity. Such result, which is
similar with the findings of previous works \cite{8}, highlights the
importance of choosing a reliable parametrization to describe
$\eta(z)$ in the early universe. Benefit from the GP technology, the
HII/QSO pairs satisfying the redshift selection criteria have a
massive growth, therefore, a considerable amount of high-redshift
samples ($z>1.4$) have been included in our analysis. Actually, such
a combination of HII regions and radio quasars enables us to get
more precise measurements at the level of $\Delta\eta\sim 10^{-3}$
by using GP reconstructed technology. Our method provided
constraints for testing validity of CDDR more stringent than other
currently available results based on real observational data.

We also consider the added benefit on the reconstruction
brought by other machine learning methods. Working on the
reconstructed luminosity and angular diameter distances with ANN, we
obtain the reconstruction of the distance duality relation $\eta(z)$
in Fig.~6, when the full data combination of HII galaxies and
compact radio quasar is considered. Similarly, the reconstructed
$\eta(z)$ function is compatible with the validity of CDDR at the
1$\sigma$ confidence level, hence there is no clear deviation from
such fundamental relation in modern cosmology. Furthermore, for both
ML approaches we find that the reconstructed errors are inconsistent
with each other, since the GP and the ANN are in principle rather
different reconstruction methods. Next, in Table 1 we show the
numerical result for CDDR parameters in the framework of four
parameterized forms. The posterior probability density functions of
CDDR parameters $\eta_j$ ($j = 0, 1 ,2$) from reconstructed HII
galaxy and radio quasar samples are shown in Fig. 7. We find that
there is some deviation from CDDR at 1$\sigma$ confidence level. The
best fitting values $\eta_0$ and $\eta_1$ are $0.031\pm0.016$ and
$0.028\pm0.014$ for first and second parameterized forms, but the
results are still consistent with zero CDDR parameters within
$2\sigma$ confidence level. Meanwhile, for the third parameterized
form, we get $\eta_2=0.031\pm0.033$ with 1$\sigma$ uncertainty.
Considering the forth form, the results are $\eta_3=0.033\pm0.043$
and $\eta_4=0.027\pm0.025$ with 1$\sigma$ errors and shown in Fig.
8. Although considering more parameters would make the constrained
precision of the CDDR parameters worse, our findings also
demonstrate the robustness of CDDR validity in two-parameter form.
In general, whatever parameterized forms are considered here, our
results indicate that there is no large extent violation of the CDDR
validity at the current observational data level, and this is one of
unambiguity conclusions in our work.

In order to highlight the potential of our method, it is necessary
to compare our results with those obtained in the previous works.
Traditionally, the angular diameter distances are derived from SZ
effect of galaxy clusters \cite{5,6}, BAOs \cite{4}, GRBs \cite{56},
and SGLs \cite{7,8}. One can combine the luminosity distance
obtained from SN Ia observations to test the validity of CDDR. Our
results are consistent with the findings of their previous works,
which confirm the validity of the CDDR at early Universe. However,
the angular diameter distances inferred from SZ effect and BAOs are
model dependent, SGLs need to the assumption of a flat Universe, and
GRBs requires additional external calibrators to calibrate it at low
redshifts. We remark here that, without any assumptions, the angular
diameter distances estimated from compact structure in radio quasars
provides a new possibility to test the fundamental relations in the
early universe model independently. More importantly, in work of
\cite{54}, they simulated gravitational wave (GW) observations based
on third-generation GW detectors Einstein Telescope and simulated
radio quasars from VLBI to test the CDDR. Their results shown that
the CDDR parameter $\Delta \eta_0 \sim0.0029$, $\Delta \eta_1
\sim0.0018$, and $\Delta \eta_2 \sim0.0051$ (at 68.3\% confidence
level) corresponding to first, second and third parameterized forms
in our work. However, we should seek other methods and technologies
until the observed GW events based on the third-generation GW
detectors will be sufficient to get statistical results in the
future.

\begin{figure}
\begin{center}
\includegraphics[width=0.9\linewidth]{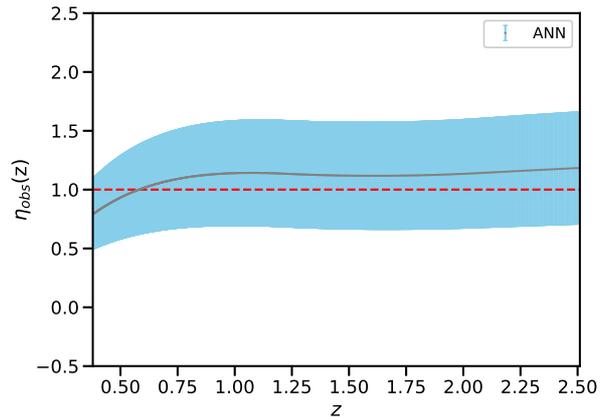}
\end{center}
\caption{Cosmic distance duality relation $\eta(z)$ for the
ANN reconstruction from the HII galaxy and compact radio quasar
sample. The dashed line at unity is $\eta(z)=1$, the solid line is
the ANN fit, and the shaded region is the $1\sigma$ ANN errors.}
\end{figure}

\begin{figure}
\begin{center}
\includegraphics[width=0.9\linewidth]{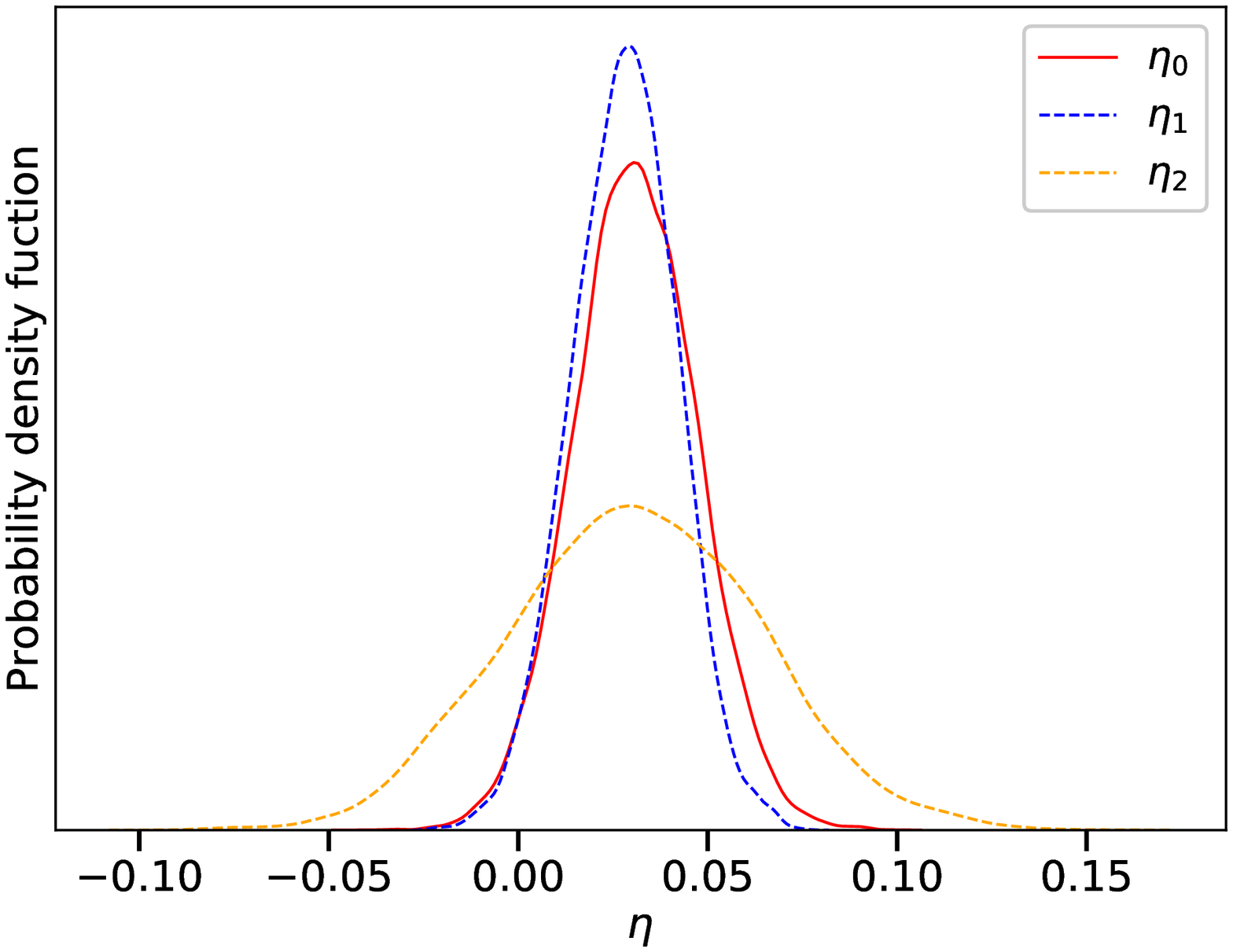}
\end{center}
\caption{The posterior probability density function with three
parameterized forms $\eta_j$ ($j = 0, 1 ,2$)  by using ANN
reconstructed HII galaxy and compact radio quasar samples.}
\end{figure}

\begin{figure}
\begin{center}
\includegraphics[width=0.8\linewidth]{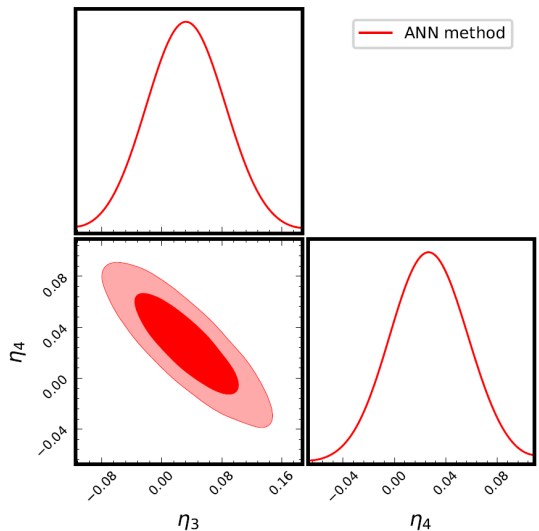}
\end{center}
\caption{ The 1D and 2D marginalized probability distributions for
the fourth CDDR parameters $\eta_3$ and $\eta_4$ by using ANN
reconstructed HII galaxy and compact radio quasar samples.}
\end{figure}

\section{Conclusion}\label{sec:con}

The cosmic distance duality relation (CDDR), as a fundamental
relation based on the metric theory of gravity, plays a important
role in modern cosmology. Possible violations of such fundamental
relation indicates that the non-conservation of the photon number
from the source to the observer due to some new physics. In this
paper, we have proposed a new model-independent method to test the
CDDR with the latest observations of HII galaxies acting as standard
candles and ultra-compact structure in radio quasars acting as
standard rulers. Specially, two machine learning reconstruction
methods, i.e., Gaussian Process (GP) and Artificial Neural Network
(ANN), are respectively applied to reconstruct the Hubble diagrams
from the observed HII galaxy and radio quasar samples. In order to
enhance the robustness of the final results, we use four commonly
used parameterized forms $\eta=1+\eta_{0}, \eta=1+{\eta_1}z,
\eta=1+\eta_{2}z/(1+z)$, and $\eta=1+\eta_{3}z+\eta_{4}z^2$ to
describe the possible violation of CDDR. Meanwhile, we also
exploit a fully agnostic reconstruction of CDDR based on two machine
learning methods, which allows us to obtain constraints without any
assumption on the redshift trend of possible deviations from CDDR.

First of all, we focus on the reconstruction of HII regions and
compact radio quasar samples through the GP method. Based on the
reconstructed standard candle and standard ruler data, we obtain the
best-fit values of the CDDR parameters $\eta_0=0.001\pm0.0031$,
$\eta_1=-0.0031\pm0.0027$, and $\eta_2=0.0014 \pm0.0054$ for the
three one-parameter forms, which are well consistent with no
violation of the cosmic distance duality relation. The results
suggest that the tests of cosmic opacity are not significantly
sensitive to the parametrization for $\eta$. For the two-parameter
parameterization, we obtain $\eta_3=-0.033\pm0.006$ and
$\eta_4=0.029\pm 0.003$ at 68.3$\%$ confidence level. A strong
degeneracy between the two redshift-dependent CDDR parameters is
also revealed in this analysis. Note that although such negative
correlation could potentially affect our test of CDDR, the validity
of such fundamental relation is still supported within $2.8\sigma$.
Therefore, our results indicate that there is no obvious violation
of the CDDR at the current observational data level, based on
Gaussian Process for the overlapping redshift domain ($z\sim 2.3$).
Moreover, we find that ultra-compact radio quasars provide an
alternative to the use of HII galaxies to confirm the validity of
the CDDR, reaching $10^{-3}$ constraints on the violation parameter.

It is still interesting to see whether those conclusions may be
changed with a different machine learning reconstruction method.
Working on the reconstructed HII regions and compact radio quasar
samples with ANN, one could derive robust constraints on the
violation parameter at the precision of $10^{-2}$, with the validity
of such distance duality relation within $2\sigma$. Although not all
of the parameterized forms support the validity of CDDR within
1$\sigma$ in the framework of GP, more convincing results are
obtained in the ANN method. Although the GP and ANN methods have
their own advantages and disadvantages \cite{50}, they both show
great potential in the studies of precision cosmology. In
the case of non-parameterized reconstruction of CDDR, our results
based on the two machine learning methods both support that there is
no obvious deviation from $\eta(z)=1$ within $1\sigma$ confidence
level. However, the statistical significance of our CDDR
reconstruction is still significantly affected by the lack of
observational data, especially at higher redshifts. Looking to the
future, an increase in the number of high-redshift standard probes
would improve the precision of our approach even further. This
strengthens our interest in observational search for more HII
galaxies and compact radio quasars with smaller statistical and
systematic uncertainties.

As a final remark, any possible deviation the CDDR might
have profound implications for the understanding of fundamental
physics and natural laws. Therefore, our results highlight the
importance of machine learning in accurately testing the current
pillars of modern cosmology and probing new physics beyond the
standard cosmological model. Summarizing, considering the wealth of
available data and various machine learning technologies in the
future, we may be optimistic to expect detecting possible deviation
from the CDDR at much higher precision.

\section*{Acknowledgments}
We thank Dr. Tian S.-X and Dr. Wang G.-J. for their helpful
discussion. This work was supported by the National Natural Science
Foundation of China under Grant Nos. 12021003, 11690023, 11633001
and 11920101003, the National Key R\&D Program of China (Grant No.
2017YFA0402600), the Beijing Talents Fund of Organization Department
of Beijing Municipal Committee of the CPC, the Strategic Priority
Research Program of the Chinese Academy of Sciences (Grant No.
XDB23000000), the Interdiscipline Research Funds of Beijing Normal
University, and the Opening Project of Key Laboratory of
Computational Astrophysics, National Astronomical Observatories,
Chinese Academy of Sciences. We also acknowledge the science
research grants from the China Manned Space Project with NO.
CMS-CSST-2021-B01.

\section*{Data Availability}

This manuscript has associated data in a data repository. The data
underlying this paper will be shared on reasonable request to the
corresponding author.

%
%

\end{document}